\documentclass[11pt]{article}

\usepackage[latin1]{inputenc}

\usepackage[T1]{fontenc}

\usepackage[bottom]{footmisc}

\usepackage[english]{babel}

\usepackage{amsmath}

\usepackage{amsfonts}

\usepackage[bottom]{footmisc}

\usepackage{amssymb}

\usepackage{mathrsfs}

\usepackage{array}

\usepackage{setspace}

\usepackage{layout}

\usepackage{graphics}

\usepackage{graphicx} 

\usepackage{bm}

\usepackage{color}

\usepackage{url} 

\usepackage{lmodern} 

\usepackage{algorithm2e}

\usepackage[top=2cm, bottom=2.5cm, left=2cm, right=2cm]{geometry} 

\newtheorem{Rem}{Remark}

\newtheorem{Theo}{Theorem}

\newtheorem{Def}{Definition}  

\newtheorem{Proof}{Proof}

\newcommand{\fonction}[5]{\begin{array}{lccl}
		#1: & #2 & \longrightarrow & #3 \\
		& #4 & \longmapsto & #5 \end{array}}

\newenvironment{psmallmatrix}{\left(\begin{smallmatrix}}{\end{smallmatrix}\right)}

\date{} 

\title{ A Hypergraph Based Approach for the 4-Constraint Satisfaction Problem Tractability}

\author{  
	$^1 $Rachid Oucheikh\footnote{The corresponding author, e-mail: rachid.oucheikh@usmba.ac.ma, B.P. 1796 Fès-Atlas, 30003 MAROC}  \hspace{1cm} $^1 $Ismail Berrada  \hspace{1cm} $^2 $Outman El Hichami \\
	\textsl{$^1$Universit\'e Sidi Mohamed Ben Abdellah},	\textsl{Faculty of Science Dhar El Mahraz, Fez, Morocco}  \\
	\textsl{$^2$Universit\'e Abdelmalek Essaâdi,}	\textsl{Faculty of Science Tetouan, Morocco}\\
	 \{oucheikh.rachid,berrada.ismail\}@usmba.ac.ma, el.hichami.outman@taalim.ma }

\begin{document}

\newpage

\maketitle 

\begin{abstract} 
	
 Constraint Satisfaction Problem (CSP) is a framework for modeling and solving a variety of real-world problems. Once the problem is expressed as a finite set of constraints, the goal is to find the variables' values satisfying them. Even though the problem is in general NP-complete, there are some approximation and practical techniques to tackle its intractability. One of the most widely used techniques is the Constraint Propagation. It  consists in explicitly excluding values or combination of values for some variables whenever they make a given subset of constraints unsatisfied. In this paper, we deal with a CSP subclass which we call 4-CSP and whose constraint network infers relations of the form: $\{  x \sim \alpha, x-y \sim \beta , (x-y) - (z-t) \sim \lambda \}$, where $x, y, z$ and $t$ are real  variables, $\alpha , \beta$ and $ \lambda $ are real constants and  $ \sim \in \{\leq , \geq \} $.  The paper provides the first graph-based proofs of the 4-CSP tractability and elaborates algorithms for 4-CSP resolution based on the positive linear dependence theory, the hypergraph closure and the constraint propagation technique. Time and space complexities of the resolution algorithms are proved to be polynomial.

\end{abstract}


\section{Introduction}

Constraint Satisfaction Problem (CSP) is a fundamental concept in constraints programing. It is fundamentally used to model and solve research problems, such as optimization, calculus and programming. CSP has received a remarkable interest over the last years, which has effectively led to the development of a rich theory that relies on techniques from various areas, especially operation research and artificial intelligence. Most real-world problems can be successfully solved using CSP; among which we can cite resource allocation, scheduling, building design, graph coloring problem, temporal reasoning, financial profits maximization, paths optimization, data clustering, tomography, and more recently natural language processing \cite{Bartak,Sally,Sqalli}.

Within the CSP framework, a problem is considered as a finite set of variables which values, satisfying certain problem-specific constraints, are assigned to. Actually, solving a CSP aims to achieve one or more of the following goals:

 	\begin{enumerate}
 		\item Finding all solutions, i.e. all combinations of values that satisfy all the constraints.
 		\item Finding one solution.
 		\item Detecting an inconsistency.
 		\item Finding an optimal solution with regard to some metrics or objective functions.
 		\item Finding all optimal solutions.
 		\item Reducing all interval domains to smaller sizes.
 		\item Reaching a solved form from which all the solutions can be easily generated.
 	\end{enumerate}

Determining whether a finite CSP (i.e. CSP with finite domain variables) has a solution is, in general, an NP-complete problem \cite{Mackworth}, which is also the case with finding one solution. An earlier attempt to solve CSPs relies on the guess and check strategy; This latter consists in guessing the assignments of all variables and checking whether they satisfy all constraints. This allows to solve  the CSP in a polynomial time. Actually, a CSP can be solved in a reasonable time either by studying the tractability of its specific subclasses or by using the heuristics and combinatorial search methods. Furthermore, the important result of Schaefer (Dichotomy Theorem) \cite{Schaefer} states that every Boolean CSP is contained in one out of six cases and gives necessary and sufficient conditions to classify the problem in polynomial-time or NP-complete. This theorem was recently generalized to a larger class of CSP (i.e. propositional logic of graphs) \cite{Bodirsky}. 
\\  

Recently, many researches have been conducted on development of effective techniques for CSPs solving, especially for the finite domain case. Examples include Constraint Propagation (CP) \cite{Apt}, Forward Checking (FC) \cite{Christian}, Maintaining Arc Consistency (MAC) \cite{Larrosa,Lecoutre}, and MAC-Backtracking techniques \cite{Zhang}. Another important topic of great application in artificial intelligence and which is considered as a special case of CSP, is the boolean SATisfiability problem (SAT) \cite{Cui,Fadi}. The SAT problem is the first known NP-complete problem, it consists in checking the satisfiability of a given propositional logic formula. Despite the SAT complexity, many of SAT instances that occur in practical issues can be solved in polynomial time. Checking the satisfiability of a formula in Conjunctive Normal Form (CNF) is a SAT subclass where each clause is limited to at most three literals (3-SAT \cite{Porschen}). It is one of Karp's 21 NP-complete problems. Besides that, 2-SAT and Disjunctive Normal Form (DNF) can be checked in linear time. 
\\

 Tractability of CSPs can be reached by considering specific classes. These classes are obtained by limiting the allowed domains or the relations which appear in constraints. For example, if the domain is binary and all variables are binary, the satisfiability is polynomial-time solvable (equivalent to 2-SAT). This paper deals with a subclass of CSPs in which constraints are expressed as inequalities written in one of the following forms:  \{ $ x \sim \alpha, x-y \sim \beta , (x-y) - (z-t) \sim \lambda $\}, such that $x, y, z$ and $t$ are real  variables, $\alpha , \beta$ and $ \lambda $  are real constants and  $ \sim \in \{\leq , \geq \} $. We denote this CSP subclass by "4-CSP". There are several reasons for which studying CSPs is important, firstly because CSP are omnipresent in a number (différent) of real-world problems and secondly for reason of their reduced complexity proven to be polynomial. Many types of hard and useful real-world problems can be modeled as 4-CSPs. The Four Phase Handshake Protocol \cite{Blunno} given in \cite{knapik}, and depicted in  Figure. 1 is one example, amongst others.  This protocol uses two clocks $x_1$, $x_2$, two parameters $minIO$, $maxIO$, and the following constraints: ($x_1 < maxIO$), ($x_1 > inIO$), ($x_2 < maxIO$), ($x_2 > inIO$), and ($(x_1-x_2) \leq (maxIO -minIO)$). Dealing with the protocol comes down to deal with its equivalent 4-CSP.  

\begin{center}
	\begin{figure} [!h]
		\centering
		\includegraphics [height = 5cm, width = 15cm]{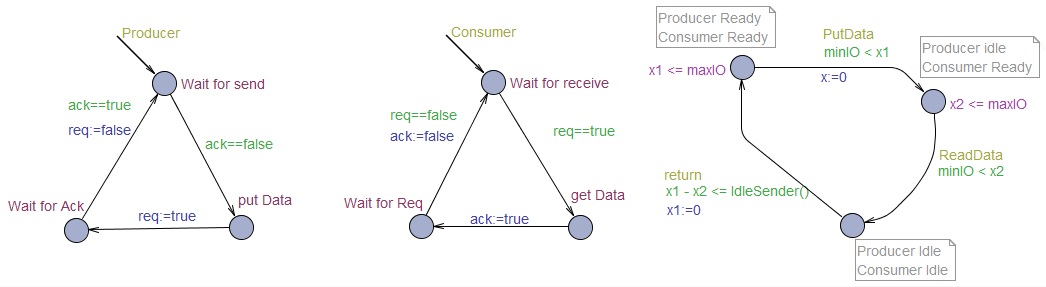}
		\label{fig:handshake}
		\caption {4-phase handshake protocol }
	\end{figure}	
\end{center}

An other relevant example that shows the utility of 4-CSPs is the verification of temporal constraints in real-time systems. Parametric Timed Automata (PTA)\cite{Alur} are among the most popular formalisms for modeling real-time systems. Almost all the systems modeled by this type of automata can be represented by a 4-CSP. PTA facilitate the manipulation of real-time systems, especially for their control and verification. Unfortunately, most of PTA verification problems are undecidable \cite{Benes,Andre}. In this model, a clock (timer) or a difference of two-clocks is compared to a linear combination of parameters. Hune et al. \cite{Hune} define the subclass "lower bound/upper bound (L/U) automata", where each parameter occurs in the timing constraints either as a lower bound or as an upper bound. In fact, L/U automata can be used to model the Fisher's mutual exclusion algorithm, the root contention protocol, and other known examples from the literature \cite{Hune}. The algorithms based on the 4-CSP framework can serve for an accelerated L/U automata verification.

In addition, the results presented in this paper show that the 4-CSP can be used to derive a numerical abstract domain \cite{Cousot,Pcousot,Putot}. 
Numerical abstract domains are widely used in static program analysis. Numerical abstraction \cite{Cousot,Dill,Mine,Claris,Halbwachs}, applied in static code analysis, provides a wider set of reachable states that guarantees the safety of the result. The challenge consists of choosing the suitable numerical abstract domain and formal methods capable to analyze all program behaviors. The abstract domain defined based on the 4-CSP extends the conjunction of octagonal invariants \cite{Mine}, called Unit Two Variable Per Inequality (UTVPI) constraints, with inequalities of the form: $\{ (x-y) - (z-t) \leq \lambda \}$. The precision of the derived domain lies between the domains of octagons \cite{Mine} and polyhedra \cite{Halbwachs}.

Using the general CSP framework to express the constraints in the aforementioned examples is very complex and almost NP-Complete. The 4-CSP framework is however precise enough to cover all their constraints and has the advantage to be linear in time and space. The 4-CSP constraint set can be seen as a subclass of the octahedron constraint set \cite{Claris} which has the form: $ \Sigma(x_i) - \Sigma(x_j) \geq k, k \in \mathbb{Q}$. However, the complexity of octahedra operations over $n$ variables is $3^n$ in memory and  $3^n$ in execution time \cite{Claris}. This is very costly compared to the complexity of our implementation proved to be cubic in the number of variables. One question that immediately comes to mind when solving the 4-CSP is why it is not enough to use the classical solving techniques, namely Linear Programming (LP). Actually, the LP seems to be not suitable for the 4-CSP presented in this paper, but it is limited just to finding an optimal solution with regard to some objective function, whereas the computation methods based on 4-CSP framework have many other goals: to guarantee the existence of a solution, to reduce all interval domains to smaller sizes and to achieve a solved form where all the solutions can easily be generated.

To sum up, the main contributions of this paper consist of:   
\begin{enumerate}
	\item Setting a theoretical basis for the  4-CSP and giving a data structure for its domains.
	
	\item Providing, based on hypergraph theory coupled with positive linear dependence theory, the first graph-based method for  the 4-CSP tractability.

	\item Developing a modified arc consistency algorithm that combines the MAC algorithm with the hypergraph closure in order to easily solve the 4-CSPs: either to check the emptiness of solution set or to list the solutions. All these operations have a polynomial time and space complexity. 
	
\end{enumerate}

The remaining of the paper is organized as follows: The second section highlights some basic definitions and provides the mathematical background of the paper. In section 3, we establish a hypergraph-based characterization of the feasibility problem. In section 4, we show the 4-CSP problem resolution methods and algorithms. Section 5 discusses the implementation issues of our approaches. Finally, section 6 concludes and draws some perspectives.

\section{Fundamental Mathematical Theories}

Throughout this paper, we use the following notations: 
\begin{itemize}
	\item $\mathbb{N}$ (resp. $\mathbb{Z}$) denotes the set of natural numbers (resp. integers) and $\mathbb{R}$ (resp. $\mathbb{Q}$) the set of real (resp. rational) numbers. 
	\item For a domain $\mathbb{T}$ ($\mathbb{R}$ or $\mathbb{Q}$), and $n \in \mathbb{N}$ : 
	\begin{itemize}
		\item $\mathbb{T}^{\geq 0}$ denotes the set $\{x \, |\, x \geq 0 , x \in \mathbb{T}\}$. $+\infty$ (resp. $-\infty$) denotes positive (resp. negative) infinity such that:
	for all $t \in \mathbb{T}$, $-\infty < t < +\infty$, $t+ (+\infty)
	= (+\infty) +t = +\infty$ and $t +(-\infty) = (-\infty) +t =
	-\infty$. $\mathbb{\overline{T}}$ denotes $\mathbb{T} \cup \{	+\infty, -\infty \}$.
	
	\item $\mathcal P \left({\mathbb{T}}\right)$ denotes the power set of $\mathbb{T}$. For a set $S\subseteq \mathcal P (\mathbb{T})$, $min(S)$ (resp. $max(S)$) is the minimal (resp. maximal) element of $S$. When $S$ has no lower bound (resp. upper bound), then $min (S)=-\infty$ (resp. $max(S)=+\infty$).
	
	\item For a set $X = \{x_{1}, x_{2}, ...., x_{n} \}$ of valued variables over $\mathbb{T}$, a valuation $\nu$ over $X$ is a function that associates to each variable of $X$, a value in $\mathbb{T}$: 
	$$\fonction{\nu}{X}{\mathbb{T}}{x_i}{\nu_i}$$
	
	$\nu$ can be seen as a vector of $\mathbb{T}^n$. $\mathcal{V}(X)$ denotes the set of valuations over $X$.
	
	 $D_{x_i}$ is the set of possible values for the variable $x_i$ and it is called domain of $x_i$. $x_0$ is a special variable that is always equal to zero i.e. $D_{x_0}=\{0\} $ and $X^0 = X \cup\{ x_0\}$. 
	
	\item $(\mathbb{T}^n, +, \times)$ denotes the n-dimensional vector space over $\mathbb{T}$. The vector $\bm{e_0} $ denotes the zero vector of $\mathbb{T}^n$.
	The set $\{ \bm{e_1},  \bm{e_2},\cdots, \bm{e_n} \}$ denotes the canonical basis (standard basis) of $\mathbb{T}^n$, that is :
	
	$$ \text{For all} \, \, \bm{e_i} = (a_j)_{j\in [1,n]} \,\Rightarrow \,
	\begin{cases}
	a_j = 1 &\text{If } j =i \\
	a_j = 0 &\text{Otherwise}
	\end{cases}
	$$
	\end{itemize}
\end{itemize}

\subsection{Positive Linear Dependence}

The theory of positive linear dependence was initiated by J. Farkas\cite{Farkas} and T. Motzkin\cite{Motzkin}, and developed by Chandler Davis \cite{Chandler}. In this paper, we consider an adaptation of this theory. Therefore, the definitions given in the rest of this section are slightly different from those of Chandler. After giving the adapted definitions, the fundamental theorem for the simple and positively dependent sets is introduced.

Let $f=(V_i)_{i\in[1,r]}$ be a family  of distinct non-empty vectors of $\mathbb{T}^n$. A \textit{strictly positive combination} of $f$ is a linear combination $\sum_{i=1}^{r} \lambda_i V_i$, with $\lambda_i \in \mathbb{N}^{>0}$. 
\begin{Def}
	
$f$ is said to be \textbf{positively independent} if none of the strictly positive combinations of $f$ is equal to $\bm{e_0}$. Otherwise, $f$ is \textbf{positively dependent} (ie. there exist some scalars $\lambda_i \in \mathbb{N}^{>0}$ such that $\sum_{i=1}^{r} \lambda_{i}V_{i}= \bm{e_0}$). 
 A positively dependent family $f$ is said to be \textbf{simple} if every subfamily $f'\subset f$ is positively independent. $\square$

\end{Def}

\begin{Theo}
	\label{theo0}
	 If $f$ is simple, then the scalars $(\lambda_i)_{i \in[1,r]} \in \mathbb{N}^{>0}$ that satisfy the equation $\sum_{i=1}^{r}  \lambda_{i}V_{i}= \bm{e_0} $ are unique (up to multiplication by a positive constant). The unique (minimal) solution is denoted by $U(f)$.  $\square$
\end{Theo}
In other words, if we have $(\lambda_i)_{i \in[1,r]} \in \mathbb{N}^{>0}$ and $(\alpha_i)_{i \in[1,r]} \in \mathbb{N}^{>0}$ such that $\sum_{i=1}^{r}  \lambda_{i}V_{i}= \sum_{i=1}^{r}  \alpha_{i}V_{i} = \bm{e_0}$, then $\frac{\alpha_i}{\lambda_i} = \frac{\alpha_j}{\lambda_j}$ for all $i,j \in [1,r]$. This can be proved based on the proof of "theorem 4.3" given by Chandler in \cite{Chandler}.


\begin{Proof} 
The proof is based on the following claim:  

\begin{itemize}
	\item For all $1\leq p < r$, the only scalars $(\alpha_i)_{i \in[1,p]} \in \mathbb{Z}$ that satisfy the equation $\sum  \alpha_{i}V_{i}= \bm{e_0} $, are $\alpha_i=0$ for $i \in [1,p]$
\end{itemize}
This claim states that any sub-family of $f$ is not  positively independent in $\mathbb{Z}$. Since $f$ is positively dependent, then there exist $(\lambda_i)_{i \in[1,r]} \in \mathbb{N}^{>0}$ such that

\begin{equation}
\label{eq1}
\lambda_1 \times V_1 + \lambda_2 \times V_2 + \cdots +  \lambda_p \times V_p + \cdots +  \lambda_r \times V_r =\bm{e_0}
\end{equation}
Now, assuming that we can find $(\alpha_i)_{i \in[1,p]} \in \mathbb{Z}$ such that $\alpha_i \neq 0$, for all $i \in [1,p]$ and: 
\begin{equation}
\label{eq2}
\alpha_1 \times V_1 + \alpha_2 \times V_2 + \cdots +  \alpha_p \times V_p =\bm{e_0}
\end{equation}
And, let $m_j = min(\{ \frac{\lambda_i}{|\alpha_i|} \, |\, i \in [1,p], \alpha_i < 0 \})$ be the minimal value reached by $\frac{\lambda_j}{|\alpha_j|}$. 

We can therefore deduce that: $$\lambda_i + m_j\times \alpha_i= \lambda_i + \frac{\lambda_j}{|\alpha_j|} \times \alpha_i = \lambda_j (\frac{\lambda_i}{\lambda_j} +  \frac{\alpha_i}{|\alpha_j|})$$
It is clear that this sum is grater that zero if $\alpha_i >0$. 

If $\alpha_i < 0$,  since 
$m_j=\frac{\lambda_j}{|\alpha_j|} \leq \frac{\lambda_i}{|\alpha_i|} \text{ implies that }  \frac{|\alpha_i|}{|\alpha_j|} \leq \frac{\lambda_i}{\lambda_j}.\text{ Thus, }  0 = ( \frac{|\alpha_i|}{|\alpha_j|} + \frac{\alpha_i}{|\alpha_j|}) \leq ( \frac{\lambda_i}{\lambda_j}+ \frac{\alpha_i}{|\alpha_j|})$
From this, we can conclude that $\lambda_i + m_j\times \alpha_i \geq 0$ for all $i \in [1,p]$ and $\lambda_j + m_j\times \alpha_j=0$. 
Finally, by multiplying equation (\ref{eq2}) with the positive scalar $m_j$ and adding it to equation (\ref{eq1}), we end up with the following new equation: 
\begin{equation}
\label{eq3}
(\lambda_1 + m_j \times \alpha_1) \times V_1 + \cdots + (\lambda_j + m_j \times \alpha_j) \times V_{j}  + \cdots + (\lambda_p + m_j \times \alpha_p) \times V_p + \lambda_{p+1} \times V_{p+1} + \cdots +  \lambda_r \times V_r =\bm{e_0}
\end{equation}
$\lambda_j + m_j\times \alpha_j=0$ means that there is a sub-family of $f$ which is positively dependent. This appears to contradict the fact that $f$ is simple.  
\\
\\
Now, if we have $(\lambda_i)_{i \in[1,r]} \in \mathbb{N}^{>0}$ and $(\alpha_i)_{i \in[1,r]} \in \mathbb{N}^{>0}$ such that $\sum_{i=1}^{r}  \lambda_{i}V_{i}= \sum_{i=1}^{r} \alpha_{i}V_{i} = \bm{e_0}$, then it is easy to see that $\alpha_r \times \sum_{i=1}^{r}  \lambda_{i}V_{i} - \lambda_r \times \sum_{i=1}^{r}  \alpha_{i}V_{i} = \sum_{i=1}^{r} ((\alpha_r\times \lambda_{i})- (\lambda_r \times  \alpha_{i}))V_{i} = \bm{e_0}$ and has at most $r-1$ vectors. From the previous result we deduce that: $\lambda_r \times \alpha_i = \alpha_r \times \lambda_i$, for all $i \leq r$. $\square$
\end{Proof}

In this way, Theorem \ref{theo0} states a fundamental result that allows the characterization of constraints to be considered while checking the emptiness of a general CSP. Furthermore, it identifies the constraints set that may have an impact on the computation of the tight bound of a given linear constraint. The case of 4-CSP is further explained in the section 2.3.

\subsection{Constraint Satisfaction Problem}

\begin{Def}[Constraint Satisfaction Problem]
	
	A Constraint Satisfaction Problem is a triplet $N=(X, D, C)$, where:
	
	\begin{itemize}
		\item $X=\{x_1,x_2,..., x_n\}$ is a set of variables.
		
		\item $D= D_{x_1}  \times D_{x_2} \times ... \times D_{x_n} $ is the domain for $X$, where $D_{x_i} \in \mathbb{T}$  is the set of possible values for the variable $x_i $. 
		
		\item $ C=\{c_1, c_2, ..., c_r\} $ is a set of constraints.

	\end{itemize}

\end{Def}	

A constraint $c_i \in C $ is a pair $<u_i, R_i> $, where $u_i \subseteq X$ is subset of $k$ variables and $R_i$ is a k-ary relation on these variables. A valuation $\nu$ satisfies $<u_i, R_i> $ if the values assigned to the variables of $u_i$ satisfy the relation $R_i$. A valuation is consistent if it verifies all the constraints in $C$ ( i.e. $\bigwedge c_i$), and is complete if it includes all variables. Each valuation that is consistent and complete is a CSP solution. By abuse of notation, $\bigwedge c_i$ denotes the CSP constraint set, and we write $C = \bigwedge c_i$.

Most of research works dealing with CSPs consider binary constraints (i.e. $k=2$). The constraints considered in this work are defined in the next paragraphs to be atomic 4-Constraints. 

\subsection{4-Constraint Satisfaction Problem}

Motivated by many real-life problems like temporal system verification,we introduce the 4-CSP with the atomic 4-constraints defined below. Let $X = \{x_{1}, x_{2}, ...., x_{n} \}$ be a set of real-valued variables over $\mathbb{T}$. 
\begin{Def}

An \textbf{atomic 4-constraint} over $X$ is an inequality of the form: 
\begin{equation*}
( \epsilon_i x_{i}- \epsilon_j x_{j}) - ( \epsilon_p x_{p}- \epsilon_q x_{q}) \sim m_{ijpq} 
\end{equation*}
where $m_{ijpq} \in \mathbb{T}$, $ \sim \in \{\leq , \geq \}$, and for all $k \in \{i,j,p,q\}$, $ \epsilon_k \in \{0,1\}$. 

An atomic 4-constraint is said to be in its \textbf{canonical form} iff for all $k \in \{i,j,p,q\}$, $\epsilon_k \neq 0$ and $"\sim"$ is equal to $"\leq"$. $\square$
\end{Def}
For instance, ($x_1 + x_2 - x_3 - x_4 \leq 4$), ($x_1 + x_2 \leq 5$), ($x_1 + x_2 - x_3 \leq 6$) and ($x_1 - x_2 - x_3 \leq 8$) are atomic 4-constraints.  It is easy to see that, by introducing a special variable $x_0$, which is always equal to zero, every atomic 4-constraint might be converted to its canonical form. For example, the 4-constraint $ (0 \times x_{i}- 1 \times x_{j}) - ( 1 \times x_{p}- 1 \times x_{q}) \sim m_{ijpq} $ can be written as: $ ( x_{0}- x_{j}) - ( x_{p}- x_{q}) \sim m_{ijpq}$.
\\
\\
The set of atomic (resp. canonical) 4-constraints over $X$ is denoted by \textbf{$\Phi(X)$} (resp. \textbf{4-$\Phi(X)$}).  In the rest of the paper, we will not distinguish between $\Phi(X)$ and 4-$\Phi(X^0)$, and we will consider only canonical 4-constraints. For a canonical 4-constraint $c_{ijpq}= (x_i-x_j)-(x_p-x_q)\leq m_{ijpq}$, we define: 
\begin{itemize}
	
	\item The \textbf{normal vector} of the hyperplane induced by $c_{ijpq}$ (variables involved in $c_{ijpq}$):  
	
	$$\fonction{F_v}{4-\Phi(X^0)}{I^{n}}{(x_i-x_j) - (x_p -x_q) \leq m_{ijpq}}{\bm{e_i - e_j - e_p +e_q}}$$

	\item The \textbf{upper bound} (\textbf{the weight function}):
	
	$$\fonction{F_b}{4-\Phi(X^0)}{\mathbb{T}}{(x_i-x_j) - (x_p -x_q) \leq m_{ijpq}}{m_{ijpq}}$$
		
	\item The \textbf{complement} :
	$$\overline{c_{ijpq}}=c_{jiqp} = ( (x_j-x_i)-(x_q-x_p)\leq m_{jiqp})$$
	Note that, $F_v(c) = -F_v(\overline{c})$ for every constraint $c\in$ 4-$\Phi(X^0)$.\\
		
\end{itemize}
\begin{Def}
	
	A \textbf{4-CSP} $S$ over $X$, is expressed as a conjunction of constraint set noted:  $$C_s = \bigwedge({(x_i-x_j)-(x_p-x_q)\leq m_{ijpq}})$$ 
	A solution of the 4-CSP is then a solution of $m$ canonical 4-constraints over $X^0$, where $m$ is the number of non-redundant conjunction terms.  $\square$ 
\end{Def}	


For a 4-CSP $S$ over $X$, we denote by $C_s$ the set of all canonical 4-constraints of $S$, and $D_s$ the domain of solutions for $4-CSP$. For a valuation $\nu \in \cal{V}$($X^{0}$), $\nu \in D_s$ iff $\nu$ satisfies all constraints of $C_s$. $D_s$ is an empty set iff for all $\nu \in \cal{V}$($X^{0}$), $\nu \not \in D_s$. As an example, the 4-CSP defined by the following 4-constraints:
$$ C_s = (x_1 + x_0 -x_2 - x_3 \leq 3) \wedge (x_2 + x_0 -x_1 - x_4\leq -4) \wedge (x_4 + x_3 - x_0 - x_0\leq 5)  \bigwedge$$ $$(x_2 +x_0 -x_0 - x_0 \leq 3) \wedge (x_3 +x_0 -x_0 - x_0 \leq 1) \wedge (x_4 +x_0 -x_0 - x_0 \leq 5) \wedge (x_1 +x_0 -x_0 - x_0 \leq 6)$$
is not empty since the valuation defined by $ (x_0,x_1,x_2,x_3,x_4)=(0,6,3,1,2) \in D_s$.
\\
\\
In order to keep bounds of constraints involved in the 4-CSP $S$, we extend the mapping $F_b$ to $C_s$, in the usual way: 

$$\fonction{F^s_{b}}{4-\Phi(X^0)}{\overline{\mathbb{T}}}{c}{
	\begin{cases}
	F_b(c) &\text{If }c \in C_s\\
	+\infty &\text{Otherwise }
	\end{cases}
}$$

In this way, $F^s_{b}$ keeps the upper bounds of constraints involved in $S$ and sets to positive infinity the other constraints not in $C_s$ (the weight function related to $S$). Finally, $S$ is said to be a \textbf{bounded 4-CSP} if there exits a scalar $w \in \mathbb{T}$ such that: 
$$ D_s \subseteq \{ \nu \, | \, \nu \in {\cal V}(X) \text{ such that} -w < \nu_i < w \}$$

\section{ Hypergraph based characterization of the tractability problem}

Graph-based algorithms has been widely used for checking the feasibility (or the emptiness) of a system of inequalities with restricted form, such as  the potential constraints conjunctions \cite{Dill} ($\bigwedge(x_i - x_j\leq m_{ij})$) and Octagons \cite{Mine} ($\bigwedge(\pm x_i \pm x_j\leq m_{ij})$). In the case of potential constraints, a data structure called Difference Bound Matrices (DBM) is used to store the system constraints. A DBM can be seen as the adjacency matrix of a directed graph $G=(N,E,w)$ (potential graph), where the set $N$ corresponds to the system variables, $E\subseteq {N}^2$ and $w\in E \mapsto\mathbb{T}$ is the weight function defined by:
$$\left\{\begin{array}{ll}
(x_i,x_j)\notin E
&\mbox{ if } {m}_{ij}=+\infty,\\
(x_i,x_j)\in E \mbox{ and }w(x_i,x_j)= {m}_{ij}\quad
&\mbox{ if } {m}_{ij}\neq +\infty\enspace.
\end{array}\right.$$
A well known result of Bellman \cite{Bellman} shows when DBMs are feasible. In fact, Bellman proves that a DBM is empty if and only if there exists, in its associated potential graph, a cycle with a strictly negative total weight. The concept of cycles (either simple cycle or closed walk) used in graph theory is able to handle constraints of the form $\pm x_i \pm x_j\leq m_{ij}$ (plan constraints). However, it will not handle constraints of the form $(x_i-x_j) -(x_p -x_q) \leq m_{ijpq}$ (hyperplane constraints). 
\\
\\
Broadly speaking, this work aims to develop scalable algorithms based on graph theory, for the feasibility checking and canonical form computation of CSPs. The question that immediately arises is can a graph theory based approach for general CSP feasibility characterization achieve similar results to that of Bellman? As will be discussed, the answer is fortunately positive for 4-CSP. This is because hypergraph theory coupled with positive linear dependence theory gives us strong theoretical tools to answer the raised question. In this paper, we restrict ourselves to 4-CSP; however, the results can be extended to CSP with constraints similar to those represented by the Octahedra abstract domain\cite{Claris}.

\begin{Def}
	A directed hypergraph \cite{Ausiello} H is a pair $(N,E)$, where $N$
	is a non empty set of nodes and $E$ is a set of hyperarcs. A hyperarc $e$ is an ordered pair $(T,h)$, with $T \subseteq N$, $T \neq \emptyset$, and $h \in N \backslash T$.  $T$ and $h$ are called the tail and
	the head of $e$, and are denoted by $tail(e)$ and $head(e)$, respectively. A weighted directed hypergraph $(N,E,w)$, is a directed hypergraph $(N,E)$ that has a positive number $w(e)$ associated with each hyperarc $e$, called the weight of hyperarc $e$. $\square$

\end{Def}
Clearly, a 4-CSP $S$ over $X$ can be easily mapped to a weighted directed hypergraph $(N,E,w)$. In fact, the set of nodes $N$ will correspond to the set of variables $X^0$. Each constraint $c_{ijpq} \in C(S)$ defines the hyperarc $e=(T,h)$ such that: $T=\{x_j,x_p\}$ and $h=\{x_i,x_q\}$. In other words, the normal vector $F_v(c_{ijpq})$ of $c_{ijpq}$, can be mapped to a unique hyperarc: positive values of $F_v(c_{ijpq})$ are mapped to the head of $e$, and negative values to the tail of $e$. 
Furthermore, we can associate a weight function to the hypergraph defined by $w(e)=F^s_b(c_{ijpq})$.  
\\
\\
Since the first papers of Berger \cite{Berger}, the hypergraph theory has been a useful tool in several fields including computer science, mathematics, bio-informatics, engineering and chemistry \cite{Zhou}. Since a hypergraph is nothing but a family of sets and for the sake of clarity, in this paper, we will use the terminology of the hypergraph theory together with the notations of positive linear dependence theory. Thus, \textbf{rather than using a hyperarc to map a 4-constraint, we use the corresponding normal vector $F_v()$} and we extend the notions of paths, cycles and minimal weights to hypergraphs in a consistent manner.  

\subsection{Hypercycles and hyperpaths}

\begin{Def}
\label{hypercycle}
Let $C = \{c_1,c_2,  \cdots , c_r\}$ be a set of distinct constraints of $\text{4-}\Phi(X^0)$. We say that $C$  generates a \textbf{hypercycle} (\textbf{h-cycle} for short) if the family $f=(F_v(c_i))_{i\in [1,r]}$ of normal vectors is positively dependent. We say that $C$  generates a \textbf{simple hypercycle } if $f=(F_v(c_i))_{i\in [1,r]}$ is simple positively dependent. $\square$
\end{Def} 
Intuitively, $C$ generates a hypercycle if we can find some strictly positive natural numbers $\lambda_i$ such that the sum $\sum \, \lambda_i F_v(c_i)$ equals the empty vector. On the one hand, this definition is quite different from those found in the literature in the sense that, the h-cycle nodes are required to appear as hyperarc tails the same number of times they appear as hyperarc heads in the associated hypergraph. On the other hand, hypercycles can be seen as a generalization of graph-based cycles where $(\lambda_i)$ are equal to 1. In fact, each edge $(x_i, x_j)$ of a cycle in a graph defines the normal vector $V_{ij}=e_i -e_j$. One can notice that $\sum \, 1\times V_{ij} = \sum \, (e_i -e_j)$  equals the zero vector. Thus, the family $(V_{ij})$ is positively dependent, which means that the set $C= \{c_1=(x_i-x_j \leq w(x_i,x_j)), c_2 = (x_j-x_l \leq w(x_j,x_l)), \cdots, c_k =  (x_k-x_i \leq w(x_k,x_i)) \}$ generates a h-cycle. Regarding the simple h-cycle, it is the hypercycle that can not be decomposed into multiple hypercycles (like elementary cycle in graphs).  Note that, the set $ \{c,\overline{c}\}$ generates a  simple h-cycle for every constraint $c \in$ 4-$\Phi(X^0)$. 
\\
\\
For instance, assuming that $X =\{x_1,x_2,x_3,x_4\}$: 
\begin{enumerate}
	\item The set $C = \{c_1 = (x_1+x_0 -x_2-x_3\leq 3), c_2=(x_2 + x_ 0- x_1-x_4\leq -4), c_3=(x_4 + x_3 -x_0 -x_0\leq 5) \}$ generates a h-cycle as $F_v(c_1)=(1,-1,-1,0)$, $F_v(c_2)=(-1,1,0,-1)$, $F_v(c_3)=(0,0,1,1)$ and $F_v(c_1)+ F_v(c_2)+ F_v(c_3)=\textbf{0}$.
	\item The set $C = \{c_1 = (x_1 + x_0 -x_2-x_3\leq 3), c_2=(x_1 + x_2 - x_3-x_0\leq -4), c_3=(x_0 + x_3 -x_1 -x_0\leq 5) \}$ generates a h-cycle as $F_v(c_1)=(1,-1,-1,0)$, $F_v(c_2)=(1,1,-1,0)$, $F_v(c_3)=(-1,0,1,0)$ and $F_v(c_1)+ F_v(c_2)+ 2\times F_v(c_3)=\textbf{0}$.
	
\end{enumerate}
In the remaining, the set of all hypercycles over 4-$\Phi(X^0)$ will be denoted:
$$ HCycle(X^0) = \{ (C,(\lambda_i)_{i\in [1,r]}) \, |\, C=\{c_1,c_2,\cdots,c_r\}, \, (\lambda_i)_{i\in [1,r]} \in \mathbb{N}^{>0}, \text{ and } \sum_{i=1}^{r} \lambda_i F_v(c_i) = \bm{e_0}\}$$
In a similar way, the notion of graph paths can be extended to \textbf{hyperpaths} as follows:

\begin{Def}
\label{hyper-path}
 Let $P = \{c_1,c_2, \cdots, c_r\} \subseteq \text{4-}\Phi(X^0)$, and $c \in \text{4-}\Phi(X^0)$. Then, $P$ generates a \textbf{hyperpath} (\textbf{h-path} for short) of $c$, if $P \cup \{\overline{c}\}$ generates a hypercycle. $P$ generates a \textbf{simple hyperpath} of $c$, if $P \cup \{\overline{c}\}$ generates a simple hypercycle. $\square$
	
\end{Def}
From the previous example, it is easy to see that $\{(x_1+x_0 -x_2-x_3\leq 3), (x_2 + x_ 0- x_1-x_4\leq -4) \}$ generates a hyperpath of  $(x_0 + x_0 -x_3 -x_4\leq 6)$. The set of all hyperpaths of $c \in \text{4-}\Phi(X^0) $ will be denoted by:
$$ HPath(c) = \{ (P,(\frac{\lambda_i}{\lambda})_{i\in [1,r]}) \, |\, P=\{c_1,c_2,\cdots,c_r\}, \, (\lambda, \lambda_i) \in \mathbb{N}^{>0}  \text{ and } F_v(\overline{c})+ \sum_{i=1}^{r} \frac{\lambda_i}{\lambda} F_v(c_i) = \bm{e_0}\}$$

\begin{Rem}
As mentioned before, each 4-constraint generates a unique hyperarc and thus the definitions \ref{hypercycle} and \ref{hyper-path} hold for the hypergraph associated to the 4-CSP.
\end{Rem}

\subsection{Some results on positive hypercycles}

Let $S$ be a 4-CSP over $X$ and $H^s=(N,E,w)$ the weighted directed hypergraph associated to $S$. As is the case with weighted graphs, $S$ defines the minimum weight hypergaph $H^s_m=(N,E,w_m)$. Before defining $H^s_m$, let us extend the weight function $F^s_b()$ (resp. $w$) of $S$ (resp. of $H^s$) to hypercycles and hyperpaths, in the usual way: 

\begin{Def}
Let $c \in 4-\phi(X^0)$ be a 4-constraint. Then: 
 \begin{itemize}
 
 \item For a h-path $(P,(\lambda_i)) \in HPath(c)$ of $c$ such that $P=\{p_1,p_2, \cdots\}$, the weight of $P$ in $S$ (and it is the same for $H^s$) is:  $w((P,(\lambda_i)))=F^s_b((P,(\lambda_i))) =  \sum \, \lambda_i F^s_b(p_i)$.
 
 \item For a h-cycle $(C,(\alpha_i)) \in HCycle(X^0)$ such that $C=\{c_1,c_2, \cdots\}$, the weight of $C$ in $S$ (the same for $H^s$) is: $w((C,(\alpha_i))) = F^s_b((C,(\alpha_i))) = \sum \, \alpha_i F^s_b(c_i)$. When $w((C,(\alpha_i)) \geq 0$, we say that $(C,(\alpha_i))$ 
 is a \textbf{positive h-cycle} of $H^s$. 
 \end{itemize}

\end{Def}
As an example, the set $C = \{(x_1+x_0 -x_2-x_3\leq 3), (x_2 + x_ 0- x_1-x_4\leq -4), (x_4 + x_3 -x_0 -x_0\leq 5) \}$ generates a positive h-cycle as the sum of these constraints is equal to $3-4+5 =4$. Next, we present some results of positive h-cycles. 

\begin{Theo}
	\label{theo2}
	Assume that all hypercycles of $H^s$ are positives and let's take $c\in 4-\Phi(X^0)$ such that $F_v(c) \neq e_0$. Then, for each h-path $P$ of $c$, we can find a simple h-path $Q$ of $c$ with a weight less than $P$. $\square$
\end{Theo}
Intuitively, the theorem establishes that we have to consider only simple h-paths when searching for the minimum weight of a hyperarc (nodes of the hyperarc).

\begin{Proof}
Let $(P,(\lambda_i)_{i\in [1,k]}) \in HPath(c)$ such that $P=\{p_1,p_2, \cdots,p_k\}$, and
\begin{equation}
	 	\label{eq4}
	F_v(\overline{c}) + \sum_{i=1}^k \lambda_i F_v(p_i) = \bm{e_0}
\end{equation}
Recall that, by definition, normal vectors of $P$ are all distinct. If $k=1$ then $P$ is simple. If $P$ is simple then $Q=P$. Now, assume that $P$ is not simple.
 Then, we can find a subset $P1=\{q_1,q_2,\cdots,q_r\}$ (at most with $k$ elements) of $P \cup \{ \overline{c}\}$ having the size $r$, such that the corresponding normal vectors are positively dependent and thus generates a h-cycle (remember that $P \cup \{ \overline{c}\}$ generates a h-cycle). 
 We identify two cases: either all $P1$ include $\overline{c}$ ($\overline{c} \in P1$) or there exists $P1$ such that $\overline{c} \not \in P1$.
 
 \begin{enumerate}
 	\item Case 1: $\overline{c}  \not \in P1$. Without loss of generality, assume that $P1=\{p_1,p_2, \cdots,p_r\}$ such that,
 	\begin{equation}
 	\label{eq5}
 	\sum_{i=1}^r \alpha_i F_v(p_i)=\bm{e_0}
 	\end{equation}
 	As all hypercycles are positive, then:
 	\begin{equation}
 	\label{eq6}
 	\sum_{i=1}^r \alpha_i F^s_b(p_i)\geq 0
 	\end{equation} 
 	
 	Let $j \leq r$ such that $m_j = \frac{\lambda_j}{\alpha_j} =min(\{ \frac{\lambda_i}{\alpha_i} \, |\, i \in [1,r] \})$. Note  that  $\frac{\lambda_i}{\alpha_i} - m_j \geq 0$, and $\frac{\lambda_j}{\alpha_j} - m_j =0$. As,   
 	
 	\begin{equation}
 	\label{eq7}
 	\sum_{i=1}^k \lambda_i F_v(p_i) =  \sum_{i=1}^r \lambda_i F_v(p_i) +\sum_{i=r+1}^k \lambda_i F_v(p_i) =\sum_{i=1}^r  m_j \times \alpha_i F_v(p_i) + \sum_{i=1}^r (\lambda_i - m_j\times \alpha_i) F_v(p_i) +\sum_{i=r+1}^k \lambda_i F_v(p_i)  
 	\end{equation}
 	\begin{equation}
 	\label{eq8}	 
 	F^s_b((P,(\lambda_i)))  =	\sum_{i=1}^k \lambda_i F_v(p_i)  = m_j \times \sum_{i=1}^r   \alpha_i F_v(p_i) + \sum_{i=1}^r \alpha_i \times(\frac{\lambda_i}{\alpha_i} - m_j) F_v(p_i) +\sum_{i=r+1}^k \lambda_i F_v(p_i)   
 	\end{equation}
 	From equations (\ref{eq8}), (\ref{eq5}) and (\ref{eq4}), we deduce that
 	\begin{equation}
 	\label{eq9}	 
 	\bm{e_0}  = F_v(\overline{c})	+ \sum_{i=1}^k \lambda_i F_v(p_i)  = F_v(\overline{c})	+ \sum_{i=1}^r \alpha_i \times(\frac{\lambda_i}{\alpha_i} - m_j) F_v(p_i) +\sum_{i=r+1}^k \lambda_i F_v(p_i)   
 	\end{equation}
 	In other words,  we have constructed a new h-path $Q$ such that  $Q \cup \{\overline{c}\}$ is a h-cycle with at most $k$ elements as $\alpha_j(\frac{\lambda_j}{\alpha_j} - m_j)=0$. Now, from equations (\ref{eq8}) and (\ref{eq6}), we deduce that $Q$ has a weight less than $P$:
 	\begin{equation}
 	\label{eq10}
 	\sum_{i=1}^r \alpha_i \times(\frac{\lambda_i}{\alpha_i} - m_j) F^s_b(p_i) +\sum_{i=r+1}^k \lambda_i F^s_b(p_i) \leq   m_j \times \sum_{i=1}^r   \alpha_i F^s_b(p_i) + \sum_{i=1}^r \alpha_i \times(\frac{\lambda_i}{\alpha_i} - m_j) F^s_b(p_i) +\sum_{i=r+1}^k \lambda_i F^s_b(p_i) 
 	\end{equation}
 	More specifically, we define $Q=(q_1,q_2, \cdots,q_k)$ and $(\beta_i)$ by: 
 	\begin{itemize}
 		\item For $i \leq r$, then
 		\begin{enumerate}
 			\item if $ \lambda_i - m_j\times \alpha_i \neq 0$ then $q_i =p_i$ and $\beta_i = \alpha_i \times(\frac{\lambda_i}{\alpha_i} - m_j)$
 			\item else drop $q_i$ from $Q$ (we drop elements of $P$ that are dependent in $P$).
 		\end{enumerate}
 		\item For $ r+1 \leq i \leq k$, then $q_i =p_i$ and $\beta_i = \lambda_i$.
 	\end{itemize}
 	From equations (\ref{eq9}) and (\ref{eq10}), it is easy to see that $Q$ is a h-path of $c$ such that: $F^s_b((Q,\beta_i) )\leq F^s_b((P,\lambda_i))$.
 	
 	Note that $Q$ has at least one element. In fact,  
 	\begin{enumerate}
 		\item if $r=k$, then there exists at least one index $i$ such that $ \lambda_i - m_j\times \alpha_i \neq 0$ otherwise $F_v(\overline{c}) = \bm{e_0}$.
 		\item if $ r<k$, at least $Q$ has $k-r\geq 1$ elements.  
 	\end{enumerate}
 	
 	In this way, we have constructed a h-path having at least one element with less weight than $P$. If $Q$ is not simple, we replace $P$ with $Q$ and repeat this reasoning until having a simple h-path of $c$.\\

 	\item Case 2: $\overline{c} \in P1$. In this case, all h-cycles of $P\cup \{\overline{c} \}$ contain $\overline{c}$. Let us show that $P$ is the sum of at least two h-paths of $c$. Without loss of generality, we assume that $P1 = \{\overline{c},p_1,p_2, \cdots,p_r\}$, with \begin{equation}
 	\label{eq11}
 	F_v(\overline{c}) + \sum_{i=1}^r \alpha_i F_v(p_i) = \bm{e_0}
 	\end{equation}
 	Note that $r <k$ as $P1 \subset P \cup \{ \overline{c} \}$. In the same way, we set $j \leq r$ such that $m_j = \frac{\lambda_j}{\alpha_j} =min(\{ \frac{\lambda_i}{\alpha_i} \, |\, i \in [1,r] \})$. First, let us show that $ m_j < 1$. In fact, if $m_j \geq 1$, then from equation (\ref{eq8}), we deduce that: 
 	\begin{equation}
 	\label{eq12}	 
 	\sum_{i=1}^k \lambda_i F_v(p_i)  = \sum_{i=1}^r   \alpha_i F_v(p_i) + (m_j-1) \times \sum_{i=1}^r   \alpha_i F_v(p_i) + \sum_{i=1}^r \alpha_i \times(\frac{\lambda_i}{\alpha_i} - m_j) F_v(p_i) +\sum_{i=r+1}^k \lambda_i F_v(p_i)   
 	\end{equation}
 	
 	From equations (\ref{eq4}), (\ref{eq11}), and by adding $F_v(\overline{c})$ to both sides of equation (\ref{eq12}), we deduce 
 	\begin{equation}
 	\label{eq13}	 
 	\bm{e_0}  =  (m_j-1) \times \sum_{i=1}^r   \alpha_i F_v(p_i) + \sum_{i=1}^r \alpha_i \times(\frac{\lambda_i}{\alpha_i} - m_j) F_v(p_i) +\sum_{i=r+1}^k \lambda_i F_v(p_i)   
 	\end{equation}

 	In other words, we found a h-cycle of $P \cup \{\overline{c}\}$ not containing $\overline{c}$ (contradiction). Thus, $m<1$. Now, let us prove that $P$ is the sum of two h-paths of $c$. Let us add $F_v(\overline{c})$ to both sides of equation (\ref{eq8}) : 
 	
 	\begin{equation}
 	\label{eq14}	 
 	F_v(\overline{c}) + \sum_{i=1}^k \lambda_i F_v(p_i)  = (1-m_j) F_v(\overline{c}) + m_j F_v(\overline{c}) + m_j \times \sum_{i=1}^r   \alpha_i F_v(p_i) +  \sum_{i=1}^r \alpha_i \times(\frac{\lambda_i}{\alpha_i} - m_j) F_v(p_i) +\sum_{i=r+1}^k \lambda_i F_v(p_i)   
 	\end{equation} 
 	
 	Again, by reducing  equation (\ref{eq14}) using equations (\ref{eq4}) and (\ref{eq11}), we have: 
 	
 	\begin{equation}
 	\label{eq15}	 
 	\bm{e_0} = (1-m_j) F_v(\overline{c}) +  \sum_{i=1}^r \alpha_i \times(\frac{\lambda_i}{\alpha_i} - m_j) F_v(p_i) +\sum_{i=r+1}^k \lambda_i F_v(p_i)   
 	\end{equation} 
 	This is nothing more than a new h-path $Q1$  of $c$ having the weight
 	\begin{equation}
 	\label{eq16}	 
 	F^s_b((Q1,(\beta_i)) = \frac{1}{1-m_j} \times (\sum_{i=1}^r \alpha_i \times(\frac{\lambda_i}{\alpha_i} - m_j) F^s_b(p_i) +\sum_{i=r+1}^k \lambda_i F^s_b(p_i)) 
 	\end{equation} 
 	In the same way, equation (\ref{eq11}) defines a h-path $Q2$ of $c$ such that:
 	\begin{equation}
 	\label{eq17}	 
 	F^s_b((Q2,(\alpha_i)) =  \sum_{i=1}^r \alpha_i F^s_b(p_i) 	\end{equation} 
 	By replacing equations (\ref{eq17}) and (\ref{eq16}) in equation (\ref{eq8}) and taking the weight function, we have: 
 	\begin{equation}
 	\label{eq18}	 
 	F^s_b((P,(\lambda_i)))  = m_j \times F^s_b((Q2,(\alpha_i)) + (1-m_j) \times F^s_b((Q1,(\beta_i))
 	\end{equation}
 	Thus, the weight of $P$ is written as an affine combination of the weights of $Q1$ and $Q2$. Thus, one of them has less weight than $P$.   Again, we found a h-path of $c$ with less weight than $P$.
 	
 \end{enumerate}
At the end, we showed that, in all cases, we can find a simple h-path having less weight than $P$. $\square$

\end{Proof}

As proved in the previous theorem, simple h-paths play an outstanding role in weighted hypergraph. In the next theorem, we establish their uniqueness.

\begin{Theo}
	\label{theo3}
		Let's assume that $P=(p_1,p_2, \cdots,p_k)$ generates in $H^s$ a simple h-path of $c\in 4-\Phi(X^0)$ such that $F_v(c) \neq e_0$. Then, there exists a unique family, noted $U(P)$, of scalars $(\lambda_i)_{i\in [1,k]}$ such that $(P,(\lambda_i)_{i\in [1,k]}) \in HPath(c)$. $\square$
		
\end{Theo} 

%
%
%
%
\begin{Proof}

Let $(P,(\frac{\lambda_i}{\lambda})_{i\in [1,k]}) \in HPath(c)$ and $(P,(\frac{\alpha_i}{\alpha})_{i\in [1,k]}) \in HPath(c)$ two simple h-paths of $c$. Thus
$$ \lambda F_v(\overline{c}) + \sum_{i=1}^{k} \lambda_i F_v(p_i) = \alpha F_v(\overline{c}) + \sum_{i=1}^{k} \alpha_i F_v(p_i) = \bm{e_o}$$
As the family $ f= \{ (F_v(p_i))_{i\in [1,k]} \cup \{F_v(\overline{c})\}$ is simple positively dependent, according to theorem \ref{theo0}, there exists a unique solution $U(f) = ((\beta_i)_{i\in [0,k]})$. Thus, there exists $m \in \mathbb{N}^{>0}$ such that $\lambda_i= m \times \beta_i$,  $\alpha_i= m \times \beta_i$, $\lambda= m \times \beta_0$ and $\alpha= m \times \beta_0$. Hence, $\frac{\lambda_i}{\lambda}= \frac{\alpha_i}{\alpha}=\frac{\beta_i}{\beta_0}$. At the end, the unique solution is $U(P)=((\frac{\beta_i}{\beta_0})_{i\in [1,k]})$. $\square$
\end{Proof}


\begin{Theo}
	\label{theo4}
	All hypercycles of $H^s$ are positive if and only if all simple hypercycles of $H^s$ are positive. $\square$
	
\end{Theo} 

\begin{Proof}
The first implication is trivial since simple hypercycles are hypercycles. Now, let $(C,(\lambda)_{i\in [1,k]})$ be a hypercycle  such that $C=\{c_1,c_2, \cdots ,c_k\}$ and let us prove that $\sum_{i=1}^{k} \lambda_i F^s_b(c_i) \geq 0$. Note that $k>1$ (a h-cycle has at least two elements), and $\sum_{i=1}^{k} \lambda_i F_v(c_i) =\bm{e_0}$ implies that $P= \{c_2,\cdots,c_k\}$ generates a h-path of $\overline{c_1}$ with weight $\sum_{i=2}^{k} \frac{\lambda_i}{\lambda_1} F^s_b(c_i)$. According to theorem \ref{theo2}, we can find a simple h-path $(Q,(\frac{\alpha_i}{\alpha})_{i\in[1,r]})$ of $\overline{c_1}$ such that $\sum_{i=1}^{r} \frac{\alpha_i}{\alpha} F^s_b(q_i) \leq \sum_{i=2}^{k} \frac{\lambda_i}{\lambda_1} F^s_b(c_i)$. As $Q \cup \{ \overline{c_1}\}$ is a simple h-cycle then $\sum_{i=1}^{r}\frac{\alpha_i}{\alpha} F^s_b(q_i) + F^s_b(c_1) \geq 0$, and thus $\lambda_1 \times (\sum_{i=2}^{k} \frac{\lambda_i}{\lambda_1} F^s_b(c_i) + F^s_b(c_1)) \geq 0$. $\square$
\end{Proof}

\subsection{Minimum weight hypergraph}

Generally, canonicity of the systems of linear inequalities is a key point when dealing with CSPs. However, as stated in \cite{Claris} (regarding roughly similar constraints i.e. octahedra constraints), computing the canonicity is hard: "finding an efficient algorithm that can compute the canonical form of an octahedron from a non-canonical system of inequalities is an open problem at the time of writing this paper ". In this part, we will introduce, for the first time, a graph-based characterization for the 4-CSP canonical form, which might lead to the development of new efficient algorithms for other CSP classes.

\begin{Def}
	Let $S$ be a 4-CSP, and $H^s=(N,E,w)$ be the weighted hypergraph associated to $S$. The minimum weight hypergraph associated to $S$ is the weighted hypergraph defined by $H^s_m=(N,E,w_m)$, where $w_m$ is the weight function defined on $C_s$ and derived from  $F^s_{bm}$ as follows:
	
	\begin{center}
	 $ w_m(c)= 
	 \left\{     
	 \begin{array}{rl}
	 F^s_{bm}(c)   & $  if  $  c \in C_s \\
	 +\infty & $  else  $
	 \end{array}\right. $	
	\end{center}
   
	Whereas $F^s_{bm}$ is defined on the set $4-\Phi(X^0)$ as follows: 		 
	$$\fonction{F^s_{bm}}{4-\Phi(X^0)}{\overline{\mathbb{T}}}{c}{min(\{ \, F^s_b((P,(\lambda_k))) \,\, | \, \, (P,(\lambda_k)) \in HPath(c) \})}\hspace{2cm} \square$$
	
\end{Def}


Since the upper bound of $c$ in $S$ might not be a tight upper bound, the minimum weight function $F^s_{bm}$ searches for the tight upper bound of $c$, if it exists, by taking the smallest bound of all h-paths of $c$.

\begin{Theo}
	\label{theo5}
	The following assertions are equivalent:
	\begin{enumerate}
		\item All simple hypercycles of $H^s$ are positive
		\item All simple hypercycles of $H^s_m$ are positive. $\hspace{2cm}\square$	
	\end{enumerate}
\end{Theo}
This theorem states that the minimum weight function preserves the positivity of hypercycles in $H^s$ and $H^s_m$. 
\begin{Proof}

	Let $C=\{c_1,c_2,\cdots,c_r\}$ be a simple h-cycle such that $\sum_{k=1}^r \, \lambda_k F_v(c_k)=\bm{e_0}$
	\begin{enumerate}
		\item Let's assume that every simple h-cycle of $H^s$ is positive and let's prove that the associated h-cycle to $C$ is positive in $H^s_m$.
		
		\begin{enumerate}
			\item  First, let us prove that for every constraint $c_k$ of $C$, $F^s_{bm}(c_k) \neq -\infty$. 
			\begin{enumerate}				
				\item According to theorem \ref{theo2}, only simple h-paths of $c_k$ can have less weight than $c_k$.  
				\item According to theorem \ref{theo3}, each simple h-path has a unique solution. 
				\item The number of combinations (subset) that we can construct from the set 4-$\Phi(X^0)$ is finite (at most $2^{n^4}$) 
			\end{enumerate}
			Thus, the set of simple h-paths of $c_k$ is finite. In other words, either $F^s_{bm}(c_k) = +\infty$ or there exists a simple h-path $P_k = (p^k_1,p^k_2,\cdots )$ of $c_k$ such that $F^s_{bm}(c_k) = \cup(P_k) = F^s_{b}((P_k,(\lambda^k_i))) = \sum_{i=1} \lambda^k_i F^s_b(p^k_i)$.     
			\item Now, for each $c_k$ of $C$, the minimal h-path of $c_k$ will be denoted by $P_k$. As $\lambda_k \times (F_v(\overline{c_k}) + \sum_{i=1} \lambda^k_i F_v(p^k_i)) = \bm{e_0}$, thus $\sum_{k=1}^r \lambda_k F_v(\overline{c_k}) + \sum_{k=1}^r \sum_{i} \lambda_k\lambda^k_i F_v(p^k_i) = \bm{e_0}$. On the one hand, we know that  $C=(c_i,c_2,\cdots,c_r)$ is a simple h-cycle such that $\sum _{k=1}^r \lambda_k F_v(c_k)=\bm{e_0}$, and we deduce that $\sum_{k=1}^r \lambda_k F_v(\overline{c_k})=\bm{e_0}$. On the other hand, $\bigcup_{k} P_k$ forms a h-cycle and thus $\sum_{k=1}^r F^s_b((P_k,(\lambda^k_i))) \geq 0$ (if simple h-cycles are positive then h-cycles are also positives, from theorem \ref{theo4}). Finally, $\sum_{k=1}^r F^s_b((P_k,(\lambda^k_i))) = \sum_{k=1}^r \sum_{i} \lambda_k\lambda^k_i F^s_b(p^k_i) = \sum_{k=1}^r \lambda_k F^s_{bm}(c_k) \geq 0$. Thus, $C$ is positives in $H^s_{m}$.

		\end{enumerate}
		\item Let's assume that every simple h-cycle of $H^s_m$ is positive and let's prove that the h-cycle associated to $C$ is positive in $H^s$. It is easy to see that the path $\{c_2,\cdots,c_r\}$ is a simple h-path of $\overline{c_1}$, and thus $F^s_{bm} (\overline{c_1}) \leq \sum_{k=2}^r \frac{\lambda_k}{\lambda_1} F^s_{b}(c_k)$. As  $F^s_{bm} (c_1) \leq F^s_{b} (c_1)$, we conclude that $0 \leq F^s_{bm} (c_1)+ F^s_{bm} (\overline{c_1}) \leq \frac{1}{\lambda_1} \sum_{k=1}^r  \lambda_k F^s_{b}(c_k)$ ($c_1$ and $\overline{c_1}$ form a simple h-cycle in $H^s_m$). $\square$. 
		
	\end{enumerate}
\end{Proof}
Next, we will give the fundamental theorem of the feasibility testing of 4-CSP.

\begin{Theo}
	\label{theo6}
	Let assume that $S$ is bounded. Then $D_s \neq \emptyset$ if and only if all simple hypercycles of $H^s_m$ are positive, where $D_s $ is the solution domain of $S$. $\square$
	
\end{Theo}
\begin{Proof} (sketch).
	
	Without loss of generality, we suppose that S is saturated, which means the existence of all constraints, and if no $c_i$ exists, we should add it (in the way that its minimal bound is infinity). Then, $H^s$ will be complete, and consequently $F^s_b$ of any constraints is bounded. 
    Assume that all simple hypercycles of $H^s_m$ are positive, and let us find a solution to $S$.
	The idea of this proof is to use the minimal function to compute minimal bounds and reduce $S$ by adding new constraints until finding a final solution. In fact, starting with $i=1$, and let us find all constraints $c_k$ such that $\{c_k,c_{i000}\}$ forms a simple h-cycle with $\lambda_k F_v(c_k) + \alpha_k F_v(c_{i000}) = \bm{e_0}$. Then, we construct a new 4-CSP $S1$ from $S$, by replacing $c_k$ with  $c'_k$ defined by: $F_v(c'_k) = F_v(c_k)$ and $F^{s1}_b(c'_k) = - \frac{\alpha_k}{\lambda_k}F^s_{mb}(c_{i000})$. In other words, we try to find the valuation $\nu \in D_s$ such that $\nu_i=F^s_{mb}(c_{i000})$. Now, $S_{1}$ defines the hypergraph $H^{s_1}$. Note that, the differences between $H^s$ and $H^{s_1}$, are only over the weight of hyperarcs (constraints) $c_k$. We can affirm then that all h-cycles of $H^{s_1}$ are positive. In fact, if we find a h-cycle which is negative, it must necessarily contain some modified constraints $c_k$: $\sum_{i=1}^{r} \lambda_i F^{s1}_b(c_i) < 0$. This is not possible because in that case we will find a new path of $c_{1000}$ strictly less than $F^s_{mb}(c_{i000})$ ($H^s_m$ is minimal). According to theorem \ref{theo4},  all h-cycles of $H^{s_1}$ are positive implies that all simple h-cycles of $H^{s_1}$ are positive. According to theorem \ref{eq5}, all simple h-cycles of $H^{s_1}_m$ will be positive. Now, given $S1$ and $H^{s_1}_m$, we restart the next iteration $i=2$, . After at most $n$ iterations, $D_s$ will be reduced to one valuation that satisfies $S$. $\square$
\end{Proof}
At the end, the minimum weight hypergraph of $S$ is saturated in the sense that all bounds are reachable.   

\begin{Theo}
\label{theo7}
If $ D_s \neq \emptyset$, then:
\begin{enumerate}
	\item For all $(i,j,p,q)$,  if $F^s_{mb}(c_{jipq}) \neq + \infty$, then there exists $\nu \in D_s$ such that $(\nu_i -\nu_j) - (\nu_p  -\nu_q) =  F^s_{mb}(c_{jipq}) $.
	\item For all $(i,j,p,q)$,  if $F^s_{mb}(c_{jipq}) = + \infty$, then for all $M < + \infty$, there exists $\nu \in D_s$ such that $(\nu_i -\nu_j) - (\nu_p  -\nu_q) \geq  M $. $\square$
\end{enumerate} 
\end{Theo}

\begin{Proof}
(similar to the previous proof)   
\end{Proof}

\section{4-Constraint Satisfaction Problems} 

As stated in the introduction, solving a given CSP aims to achieve one or more goals. In the case of our 4-CSP, we aim to: 
\\
\begin{itemize} 
\item[\textbullet] Detect an inconsistency.
\item[\textbullet] Guarantee the existence of at least one solution.
\item[\textbullet] Reduce all interval domains to smaller sizes.
\item[\textbullet] Achieve a solved (or canonical) form wherefrom all solutions can be generated easily.
\\
\end{itemize}

As will be detailed in this section, these goals can be achieved using the hypergraph-based characterization introduced in the previous section. 
\\

Computing the canonical form of a 4-CSP, using the minimal weight function, will provide a useful mechanism to solve many problems modeled by 4-CSP. However, finding an efficient algorithm that can compute the minimal weight function is, in the general case, an open problem at the time of writing this paper. Note that, computing the minimal weight by finding all $HPath$, is a hard problem since there are exponential number of $HPath$. Thus, as long as an upper approximation can be guaranteed, an exact representation of a 4-CSP is not required. Keeping this fact in mind, we introduce some fundamental results that will allow us to compute either the canonical form (for some special cases), or an upper approximations of the canonical form. The next theorem gives the necessary conditions to be verified by the minimal weight function. 
\begin{Theo}
	\label{theo8}
	Let  $x_i, x_j, x_p, x_q, x_k, x_l$ be six variables of $X^0$ and let $M_{ijpq}$ denotes the minimal bound $F^s_{bm}(c_{ijpq})$ of a constraint $c_{ijpq} = ((x_i-x_j) - (x_p - x_q) \leq m_{ijpq})$. Then,
	\begin{enumerate}
		\item $M_{ijpq}=M_{qpji}=M_{ipjq}$
		\item $M_{ijkk} = M_{ij00}$
		\item $M_{ijji} = 2M_{ij00}$ 
		\item $M_{ijpq} \leq M_{ijkl} + M_{klpq}$
		\item $M_{ijpq} \leq M_{iklq} + M_{kjpl}$ $\square$ 
		
	\end{enumerate} 
	
\end{Theo}

\begin{Proof}
	
	The proof of the first point is based on the fact that $F_v(c_{ijpq})= F_v(c_{ipjq}) = F_v(c_{qpji})$ and thus  $HPath(c_{ijpq})= HPath(c_{ipjq}) = HPath(c_{qpji})$. 
	The same remark holds for points 2 and 3: $F_v(c_{ijkk})=  F_v(c_{ij00})$ and $F_v(c_{ijji})=2F_v(c_{ij00})$. 
	\\
	Now, the idea of proving the 4 point comes from the fact that $M_{ijpq}$ is either $-\infty$ (presence of negative hypercycles in $S$) or reached by a h-path. Here we give only the proof for the first point; the last one can be proved similarly.
	\begin{itemize}
		\item Let assume that $M_{ijkl} \neq -\infty$ and $M_{klpq} \neq -\infty$. Then, there exist two h-paths $(P1,(\lambda_i)) \in HPath(c_{ijkl})$ and $(P2,(\lambda'_i)) \in HPath(c_{klpq})$ such that:
		\begin{enumerate}
			\item  $P1=\{c_1,c_2,\cdots\}$, $M_{ijkl} = F^s_b((P1,(\lambda_i)))=\sum \lambda_i F_v(c_i) $, and $F_v(\overline{c_{ijkl}}) + \sum \lambda_i F_v(c_i) = \bm{e_0}$
			\item  $P2=\{c'_1,c'_2,\cdots\}$, $M_{klpq} = F^s_b((P2,(\lambda'_i)))=\sum \lambda'_i F_v(c'_i) $, and $F_v(\overline{c_{klpq}}) + \sum \lambda'_i F_v(c'_i) = \bm{e_0}$. 
		\end{enumerate}
		
		Since $F_v(\overline{c_{ijkl}}) + F_v(\overline{c_{klpq}}) = F_v(\overline{c_{ijpq}})$,  and $F_v(\overline{c_{ijpq}}) + \sum \lambda_i F^s_b(c_i) + \sum \lambda'_i F^s_b(c'_i) = \bm{e_0}$,	$P_1\cup P_2$ generates a h-path of $c_{ijpq}$. 
		Then, $M_{ijpq}$ is less than the h-path bound associated to $P_1\cup P_2$ which has as bound of $ \sum \lambda_i F^s_b(c_i) + \sum \lambda'_i F^s_b(c'_i) = M_{ijkl}+M_{klpq}$ and thus $M_{ijpq}\leq M_{ijkl}+M_{klpq}$.
		
		\item Now, assume that $M_{ijkl} = -\infty$. As every h-path of $c_{ijkl}$, on the one side, can be extended to a h-path of $c_{ijpq}$, and on the other side, has a new h-path smaller than it ($M_{ijkl} = -\infty$), then $M_{ijpq} = - \infty$. A similar proof remains valid if $M_{klpq} = -\infty$ $\Box$
		
	\end{itemize}
\end{Proof}

The next theorem, presented below, deals with 4-CSP subclasses solutions.

\begin{Theo}
\label{theo9}
	The way we can get the canonical form is given for some subclasses of 4-CSP as follows:
	\begin{enumerate}
	\item \textbf{Octagon forms}: if all 4-constraints are of the form $ (\pm x_i \pm x_j \leq k)$, then the canonical form is given by the first four points of theorem \ref{theo8}.
	
	\item \textbf{Upper bound forms}: if all 4-constraints are of the form $( x_i - x_j \leq x_p + k)$, then the canonical form is given by the five points of theorem \ref{theo8}.

	\item \textbf{Lower bound forms}: if all 4-constraints are of the form $(   x_p \leq x_i - x_j + k)$, then the canonical form is given by the five points of theorem \ref{theo8}.
	 	
	\end{enumerate} 
	
\end{Theo}

\begin{Proof}
	
	The different subclasses of 4-CSP are based on the nature of their constraints, which result in the three following subclasses:
	
	\begin{enumerate}
		
		\item Octagon subclass:
		
		Foremost, the octagon inequalities are translated into the 4-$CSP$ atomic constraints as follows:	$x_i-x_j-x_0+x_0 \leq M_{ij00} $, $x_0-x_i-x_j+x_0 \leq M_{0ij0} $, and so on. The initial constraints are then: $c_{i000}, c_{ij00}, c_{0ij0}, c_{00ij}$. If we take into account the four first points of Theorem 8, all other constraints can be derived from these initial ones. $c_{ijpq} $ can be obtained, for instance, from $c_{ij00}$ and $c_{00pq}$.
		\\
		 The challenge is to prove that if all the minimal bounds of the octagon verify the first four points of Theorem 8, then the octagon is surely in its canonical form. Formally speaking:

	\begin{center}
			$\forall (x_i,x_j) \in \mathbb{R}^2, \pm x_i \pm x_j \leq k \implies \nexists k' \in \mathbb{R}$ such that: $ \pm x_i \pm x_j \leq k' \leq k $.
	\end{center}
		

This assertion will be proved by contraposition. Suppose that the octagon is not canonic even when the four points of Theorem 8 are verified, then  there exists, for instance, a hyperpath $P=(p_1, p_2, \ldots , p_n)$ such that: $ F_v(P)= F_v(c_{ij00})$ and $ F^s_{mb}(P) < M_{ij00}$. $P \cup c_{ij00} $  generates a h-cycle, which means: $ F_v(c_{ji00}) + \Sigma^n_{i=1} \lambda_{i}F_v(p_i)= e_0 $.
		
		Suppose that this hyperpath length equals to one, i.e. $P=(p_1, p_2)$ then: $ \exists M_{i000}$ and $M_{0j00} $ such that: $M_{i000}+M_{0j00} < M_{ij00} $. This is absurd since the fourth point of Theorem 8 is already fulfilled.
		Now, for any hyperpath length, i.e. $P=(p_1, p_2, \ldots , p_n)$ then $\exists M_{c_1}, M_{c_2}, ...,M_{c_n} $ such that: $ \sum M_{c_i} < M_{ij00}$. Each constraint $c_i$ is either in or derived from the initial form. Let's replace all constraints by their initial ones, for example: $c_{ijpq} $ can be replaced by two constraints having the lowest bounds ( $c_{ij00} $ and $c_{00pq}$ for instance ). By doing this we can deduce by induction that: $\exists M_{c_n}$ and $ M_{c_m}$ such that: $M_{c_n}+M_{c_m} < M_{ij00} $ and this contradicts the fourth point of Theorem 8.

	 \item The upper bound subclass:
	 
	  This form contains three variables per inequality. Considering the proof of the octagon case and Theorem \ref{theo8}, the need for using the first four points of the Theorem \ref{theo8} to get the canonical form can be proved easily. Let us prove the necessity of the fifth point: $M_{ijp0}\leq M_{ikl0}+M_{kjpl}$:
	 
	   Let assume that $M_{ikl0} \neq -\infty$ and $M_{kjpl} \neq -\infty$. Then, there exists two h-paths $(P1,(\lambda_i)) \in HPath(c_{ikl0})$ and $(P2,(\lambda'_i)) \in HPath(c_{kjpl})$ such that:
	       	\begin{enumerate}
	       		\item  $P1=\{c_1,c_2,\cdots\}$, $M_{ikl0} = F^s_b((P1,(\lambda_i)))=\sum \lambda_i F_v(c_i) $, and $F_v(\overline{c_{ikl0}}) + \sum \lambda_i F_v(c_i) = \bm{e_0}$
	       		\item  $P2=\{c'_1,c'_2,\cdots\}$, $M_{kjpl} = F^s_b((P2,(\lambda'_i)))=\sum \lambda'_i F_v(c'_i) $, and $F_v(\overline{c_{kjpl}}) + \sum \lambda'_i F_v(c'_i) = \bm{e_0}$. 
	       	\end{enumerate}
	       	
	       	As $F_v(\overline{c_{ikl0}}) + F_v(\overline{c_{kjpl}}) = F_v(\overline{c_{ijp0}})$,  and $F_v(\overline{c_{ijp0}}) + \sum \lambda_i F^s_b(c_i) + \sum \lambda'_i F^s_b(c'_i) = \bm{e_0}$, thus
	       	$P_1\cup P_2$ generates a h-path of $c_{ijp0}$. 
	       	Finally, $M_{ijp0}$ is less than the h-path bound associated to $P_1\cup P_2$ which has as bound of $ \sum \lambda_i F^s_b(c_i) + \sum \lambda'_i F^s_b(c'_i) = M_{ikl0}+M_{kjpl}$ and thus $M_{ijp0}\leq M_{ikl0}+M_{kjpl}$,  which is a special case for the fifth property: $M_{ijpq} \leq M_{iklq} + M_{kjpl}$.
	       	
	    \item The lower bound subclass:
	    
	     This form contains also three variables per inequality. The result is proved in the same manner as upper bound forms taking into account just the order matter.$\Box$
	
	\end{enumerate}

\end{Proof}

\section{Implementation}

After presenting all necessary ingredients and theoretical backgrounds related to the 4-constraint satisfaction problem, we discuss in this section the implementation of a 4-CSP
detail, from an implementation point of view, how 4-CSP can be stored and how efficient algorithms can be developed for computing canonical forms and testing the emptiness of a 4-CSP.

\subsection{2D-DBM data-structure}

Difference Bound Matrix (DBM) is a square matrix $M$ where each coordinate $m_{kl}$ represents the upper bound of the difference $x_l - x_k$. For example, the following constraints $ x_{1} \leq 4$ (equivalent to $x_{1} - x_{0} \leq 4$), $x_{2} \leq 3, x_{2} \geq 5 $, $ 8 \geq x_{2} - x_{1} \geq 6 $ can be represented by the following DBM:

\begin{center}

$
 M= \bordermatrix{ & x_{0} & x_{1} & x_{2}  \cr
 x_{0}  & 0  & 4  & 3  \cr
 x_{1}  &  0 & 0  & 8  \cr  
 x_{2}  &  -5 & -6 & 0  \cr 
 }
$\\
\end{center}

To implement and facilitate the manipulation of the 4-CSP domains, a suitable data structure is needed. Therefore, DBM is extended in two dimensions to obtain the so-called "2D-DBM".
A 2-Dimensions Difference Bound Matrix (2D-DBM) is a square matrix $M$ where $m_{kl}$ is the upper bound $M_{ijpq}$ of the constraints $C_{ijpq}$, for $ 1 \leq k,l \leq (n+1)^{2}$:  lines and columns become difference of variables instead of variables, as depicted in Figure \ref{fig:2D-DBM}.  

\begin{figure}[!h]
\begin{center}

$
 M= \bordermatrix{ & x_{0}-x_{0} & x_{0}-x_{1} & \hdots & x_{i}- x_{j} & \hdots & x_{n}- x_{n} \cr
 x_{0}-x_{0} & 0  & M_{0100}  &  \hdots & M_{ij00} & \hdots   & M_{nn00} \cr
 x_{0}-x_{1}  &  M_{0001} & M_{0101}  & \hdots & M_{ij01} & \vdots & M_{nn01}\cr 
            &  \vdots & \vdots  & \ddots    & \vdots & \ddots & \vdots   \cr 
 x_{p}-x_{q}  &  M_{00pq} & M_{01pq}  & \hdots & M_{ijpq} & \vdots   &  M_{nnpq} \cr
						&  \vdots & \vdots  & \ddots  & \vdots  & \ddots & \vdots  \cr 
 x_{n}-x_{n}  &  M_{00nn} & M_{01nn} & \hdots & M_{ijnn} & \hdots   & M_{nnnn}\cr 
 }
$
\end{center}

\caption{2D-BDM data structure.}
\label{fig:2D-DBM}
\end{figure}

\subsection{Canonical form computation algorithm}

The idea of computing the canonical form (or, sometimes, just  an upper approximation) of a given 2D-DBM is based on Theorems \ref{theo8} and \ref{theo9}. In fact, Theorem \ref{theo8} gives the necessary conditions to be fulfilled by any canonical 4-CSP (e.g canonical 2D-DBM), whereas Theorem \ref{theo9} establishes special cases where some of these conditions are sufficient. From the viewpoint of graph theory, both theories rely on the minimal weight hypergraph (the hypergraph closure) associated to a 2D-DBM.

\subsubsection{The hypergraph closure}

The first algorithm developed in this paper, for computing the canonical (or, sometimes, just an upper approximation) form of a given 2D-DBM and testing the emptiness of the solution set, is illustrated in the Algorithm \ref{algo_disjdecomp}. From the viewpoint of graph theory, Algorithm \ref{algo_disjdecomp} allows to minimize the hypergraph associated to a given  4-CSP and to check the existence of a negative hypercycle.

\begin{algorithm}
\SetKwData{Left}{left}
\SetKwData{This}{this}
\SetKwData{Up}{up}
\SetKwFunction{Union}{Union}
\SetKwFunction{FindCompress}{FindCompress}
\SetKwRepeat{Do}{do}{while}
\SetKwInOut{Input}{input}
\SetKwInOut{Output}{output}
\Input{2D-DBM}
\Output{Canonical 2D-DBM (or upper approximation of the canonical form in the worst case)}
\BlankLine
\BlankLine


 \Do{2D-DBM is not yet stationary}{
          \ForEach{cell $M_{ijpq}$ in 2D-DBM representing a 4-CSP constraint}{
	$M_{ijpq} := min(M_{ijpq}, M_{ijkl} + M_{klpq}, M_{iklq} + M_{kjpl})$ 
	}
	
	\BlankLine

	Update cells in order to ensure the following equalities:
	 
	\hspace {0.5cm} {$M_{ijpq}:=M_{qpji}:=M_{ipjq}$}
	
	\hspace {0.5cm} {$M_{ijkk}:=M_{ij00}$}

	\hspace {0.5cm} {$M_{ijji}:=2M_{ij00}$}

	\BlankLine

}

\BlankLine
\BlankLine

\caption{Skeleton of the hypergraph closure}
\label{algo_disjdecomp}
\end{algorithm}

A solution is guaranteed if the diagonal of the final canonical 2D-DBM does not contain any negative cell (i.e. there is no negative hypercycle in the hypergraph). Thus, we obtain at least one solution: the variable valuations contained in the first column. Note that each iteration of the algorithm presents the constraint propagation technique, since the changing of one constraint upper bound impact the upper bounds of the others. In fact, each iteration strengthens the bounds of each system constraint, which means that it excludes quickly many values from the variables domains. Consequently, the constraint propagation process is accelerated.

\subsubsection{From the hypergraph closure to the 4-CSP tractability}

\begin{algorithm}
	\SetKwData{Left}{left}
	\SetKwData{This}{this}
	\SetKwData{Up}{up}
	\SetKwFunction{Union}{Union}
	\SetKwFunction{FindCompress}{FindCompress}
	\SetKwRepeat{Do}{do}{while}
	\SetKwInOut{Input}{input}
	\SetKwInOut{Output}{output}
	\SetKw{KwTo}{to}
	\SetKwComment{tcc}{/*}{*/}
	\Input{2D-DBM}
	\Output{Canonical 2D-DBM }
	\BlankLine	
	\BlankLine	
Variables
i, j, p, q, k, l, s: Integers\; 

/* In this algorithm $[i] $ denotes the integer value of $i$ and $n+1$ the number of domain variables (including the $x_0$ variable which is always null), and we note:

	\begin{itemize} 
		
		\item[ $\bullet$ ]  $i = [l/(n+1)] $
		\item[ $\bullet$ ]  $j= l-[l/(n+1)]*(n+1)$
		\item[ $\bullet$ ] $p= [k/(n+1)] $
		\item[ $\bullet$ ] $q= k-[k/(n+1)]*(n+1)$ */
		
	\end{itemize}    

M: Table\; 

$iter:=1$;

\Do{ $iter <= pow((n+1), 4)/2$ }  {
\For{$k=1$ \KwTo $pow((n+1), 2)$ }{   
\For{$l=1$ \KwTo $pow((n+1), 2)$ } {

\tcc{The following loop serves to update the matrix in order to verify the two last points of the Theorem 8.}

\For{  $s=1$ \KwTo  $pow((n+1), 2)$ } {
$M[k,l] := min( M[k,l], M[k,s] + M[s,l],$ \
$M[p*(n+1)+i , s] + M[s , q*(n+1)+j ], $ \
$M[p*(n+1)+q , s] + M[s , i*(n+1)+j ], $ \
$M[j*(n+1)+q , s] + M[s , i*(n+1)+p ], $ \
$M[(s/[n+1])*(n+1)+i , q*(n+1)+s-[s/(n+1)]*(n+1) ] $ \
$+ M[j*(n+1)+s/[n+1] , (s-[s/(n+1)]*(n+1))*(n+1)+p ] ) $;
}

\tcc{The following loop serves to update the matrix in order to verify the three first points of the Theorem 8.}

\For{ $p=1$ \KwTo  $pow((n+1), 2)$} {
$M[k,l] := min(M[k,l], M[(p*(n+1)+q , i*(n+1)+j ], M[p*(n+1)+i , q*(n+1)+j] )$;

	\BlankLine
	\BlankLine
	
\If(\tcc*[h] { $ M_{ijkk} = M_{ij00} $}) {$([l/(n+1)] = l-[l/(n+1)])$}  {$M[k,l] := M[k,0]$;}

\If(\tcc*[h]{ $M_{ijji} = 2M_{ij00}$ }) {$([k/(n+1)] = l-[l/(n+1)]$ and $k-[k/(n+1)] = [l/(n+1)])$}  {$M[k,l]:=2*M[k,0]$; }    
}
}
} 
$iter++$; 
} 
			     
	\BlankLine
       
\caption{Canonical form of a 2D-DBM}
\label{algo_closure}
\end{algorithm}

For the sake of clarity, we presented in Algorithm \ref{algo_disjdecomp} only the skeleton of the hypergraph closure, without giving technical details about how operations will be implemented or when the algorithm will terminate. Technically, in the implementation, we use two two-dimensional tables, with $ (n+1)^2 $ columns and $ (n+1)^2 $ lines. The 2D-DBM is rewritten in a way that: column $C_{ij}$ (resp. line $L_{ij}$) of 2D-DBM which represents the variable $x_i-x_j$ becomes the column $C_{i*n+j} $ (resp. line  $L_{i*n+j} $).

\paragraph{The algorithm complexity analysis.} It is obvious that the canonical form is obtained in at most $(n+1)^4/2 $ iterations. This maximum number of  iterations is achieved if we suppose that just two difference variables are related pairwise, that way we will have $(n+1)^4/2 $ binary classes. \\

Our algorithm is polynomial in time and space. In fact it has a complexity $ \mathcal{O}(n^{10})$. This reflects the efficiency of our algorithm compared with the other approximation algorithms that infer complex constraints, and the precision obtained besides using just binary constraints interested in by the majority of works.

\subsubsection{ The whole algorithm }

Up to now, the domain of solutions is very reduced, it remains fair to extract the solution combinations. For this, we add to our algorithm the last version of the Arc Consistency algorithm AC2001 \cite{Bessiere}. AC2001 takes as an only input the Constraint Network resulted from the hypergraph closure algorithm, which is very reduced. Therefore, it provides the set of variable values quickly.

\begin{algorithm}
	\SetKwData{Left}{left}
	\SetKwData{This}{this}
	\SetKwData{Up}{up}
	\SetKwFunction{Union}{Union}
	\SetKwFunction{FindCompress}{FindCompress}
	\SetKwRepeat{Do}{do}{while}
	\SetKwInOut{Input}{input}
	\SetKwInOut{Output}{output}
	\SetKw{KwTo}{to}
	\SetKwComment{tcc}{/*}{*/}
	\Input{Constraint Network $(N,D,C)$}
	\Output{Solutions of the CN }
	
	\BlankLine
	\BlankLine

\hspace{0.6cm} 1  \hspace{0.8cm}  The hypergraph closure function on 2D-DBM (Algorithm \ref{algo_closure})

\hspace{0.6cm} 2  \hspace{0.8cm}  Take bounds of variables (for domains $D$) and bounds of binary relations from 

\hspace{1.9cm} 2D-DBM (for constraints $C$)

\hspace{0.6cm} 3  \hspace{0.8cm}  Accomplish the arc consistency algorithm AC2001

     \BlankLine
	 \BlankLine
	 		     
    \caption{The Whole Algorithm Skeleton}
    \label{algo_whole}
 \end{algorithm}

\section{Conclusion}

In this paper, we have introduced a subclass of CSP named 4-CSP. As it has been shown, studying 4-CSP can be of great importance, considering their omnipresence in many real problems as well as their reduced complexity proven to be polynomial. In comparison with the other variants of CSP, the 4-CSP is more rich than binary CSP in terms of invariants precision; and less complex than the general CSP in terms of implementation cost, since it is proved to be cubic in the number of system variables. The main contribution of this paper consists of providing a complete framework for the 4-CSP, including the theoretical background and the implementation issues. 

In addition, we have also provided the first answer, to the best of our knowledge, to the following fundamental problem : can we build a scalable and graph theory based algorithms for CSP tractability similar to those of Bellman? Thanks to the hypergraph theory coupled with positive linear dependence theory, a positive answer has been proved for the 4-CSP class. This result might be extended to CSP with constraints similar to those of the Octahedra \cite{Claris}.

Finally, in order to represent and manipulate 4-CSP, we have defined a suitable data-structure called 2D-DBM, and elaborated the algorithm able to obtain the canonical form for this structure.


\begin{thebibliography}{8}


\bibitem{Allender} E. Allender, M. Bauland, N. Immerman, H. Schnoor, H. Vollmer. The complexity of satisfiability problems: Refining Schaefer's theorem. Journal of Computer and System Sciences, Volume 75, Issue 4, 2009, pp. 245-254, ISSN 0022-0000.


\bibitem{Fadi} F. A. Aloul, Search techniques for SAT-based Boolean optimization, Journal of the Franklin Institute, Volume 343, Issues 4-5, July-August 2006, pp. 436-447. 

\bibitem{Alur} R. Alur, T.A. Henzinger, and M.Y. Vardi.  Parametric real-time reasoning.  In Proceedings of 25th Annual Symposium on Theory of Computing (STOC-93), pp. 592-601. ACM Press, 1993. 

\bibitem{Andre} É. André, and N.  Markey. Language preservation problems in parametric timed automata. In Formal Modeling and Analysis of Timed Systems, pp. 27-43. Springer International Publishing, 2015. 


\bibitem{Apt} K. R. Apt. The essence of constraint propagation. Theoretical Computer Science, 221(1-2), pp. 179-210, 1999.

\bibitem{Ausiello} G. Ausiello, P. G. Franciosa, and D. Frigioni. Directed hypergraphs: Problems, algorithmic results, and a novel decremental approach. Theoretical Computer Science. Springer Berlin Heidelberg, pp. 312-328, 2001. 

\bibitem{Bartak} R. Barták, Miguel A. Salido, and F. Rossi. New trends in constraint satisfaction, planning, and scheduling: a survey. Knowledge Eng. Review, 25(3), pp. 249-279, 2010.


\bibitem{Bellman} R. Bellman. On a routing problem. In Quarterly of Applied Mathematics, Vol. 16. pp. 87-90, 1958.


\bibitem{Berger} C. Berge. Graphes et Hypergraphes, Dunod, Collection Monographies Universitaires de Mathématiques n.37, janvier 1970.

\bibitem{Benes} N. Benes, P. Bezdek, K. G. Larsen, and J. Srba. Language Emptiness of Continuous-Time Parametric Timed Automata. arXiv:1504.07838., 2015.


\bibitem{Christian} C. Bessière, P. Meseguer, E. C. Freuder, Javier Larrosa, On forward checking for non-binary constraint satisfaction, Artificial Intelligence, Volume 141, Issue 1, 2002, pp. 205-224, ISSN 0004-3702. 

\bibitem{Bessiere}  C. Bessière, J. C. Régin, R. H. C. Yap, Yuanlin Zhang, An optimal coarse-grained arc consistency algorithm, Artificial Intelligence, v.165 n.2, pp. 165-185, July 2005.

\bibitem{Blanchet}  B. Blanchet, P. Cousot, R. Cousot, J. Feret, L. Mauborgne, A. Miné, D. Monniaux, and X. Rival. A static analyzer for large safety-critical software. In PLDI, pp. 196-207. ACM Press, 2003.


\bibitem{Blunno} I. Blunno, J. Cortadella, A. Kondratyev, L. Lavagno, K. Lwin, C.P Sotiriou. Handshake protocols for de-synchronization. In Proc. of 10th International Symposium on Advanced Research in Asynchronous Circuits and Systems (ASYNC 2004), pp. 149-158, (2004).


\bibitem{Bodirsky} M. Bodirsky and M. Pinsker. Schaefer's theorem for graphs. Journal of the ACM, Vol. 62(3), Article 19, pp. 1-52. 2015.


\bibitem{Bozzeli} L. Bozzelli and L. Salvatore. Decision problems for lower/upper bound parametric timed automata. Formal Methods in System Design, 2009, vol. 35, no 2, pp. 121-151.

\bibitem{Sally}  S. C. Brailsford, C. N. Potts, Barbara M. Smith. Constraint satisfaction problems: Algorithms and applications, European Journal of Operational Research, Volume 119, Issue 3, 16 December 1999, pp. 557-581, ISSN 0377-2217.

\bibitem{Chandler} D. Chandler. Theory of positive linear dependence. American Journal of Mathematics, 1954, pp. 733-746.


\bibitem{Chen} L. Chen, A. Miné, J. Wang, and P. Cousot. Interval polyhedra: An abstract domain to infer interval linear relationships. International Static Analysis Symposium. LNCS, vol. 5673, pp. 309-325. Springer, 2009.



\bibitem{Wang} L. Chen, J. Liu, A. Miné, D. Kapur and J. Wang. An Abstract Domain to Infer Octagonal Constraints with Absolute Value. International Static Analysis Symposium. LNCS, vol. 8373, pp. 101-117. Springer, 2014.


\bibitem{Claris}  R. Claris\`{o}, J. Cortadella. The Octahedron Abstract Domain. In Science of Computer Programming, 64(2007):115-139. 


\bibitem{Cousot} P. Cousot, R. Cousot. Abstract interpretation: a unified lattice model for static analysis of programs by construction or approximation of fixpoints. In Symposium on Principles of Programming Languages, pp. 238-252. ACM Press (1977).

\bibitem{Pcousot} P. Cousot and R. Cousot. Abstract Interpretation and Application to Logic Programs. Journal of Logic Programming 13(2-3), pp. 103-179. 1992.

\bibitem{Cui}  T. Cui, F. Franchetti. Autotuning a Random Walk Boolean Satisfiability Solver. Procedia Computer Science, Volume 4, 2011, pp. 2176-2185, ISSN 1877-0509.

\bibitem{Dill} D. L. Dill. Timing assumptions and verification of finite-state concurrent systems. In Automatic Verification Methods for Finite State Systems, LNCS 407, pp. 197-212. Springer-Verlag, 1989.

\bibitem{Farkas} J. Farkas. Uber die Theorie der Einfachen Ungleichungen. Journal für die Reine und Angewandte Mathematik, Vol. 124, pp. 1-24, ISSN:0075-4102. 1902. 

\bibitem{Hune} T. Hune, T. Romijn, M. Stoelinga, and F Vaandrager. Linear parametric model checking of timed automata. J Log Algebraic Program 52-53:183-220, 2002.

\bibitem{Halbwachs} N. Halbwachs, Y.-E. Proy, and P. Roumanoff. Verification of real-time systems using linear relation analysis. Formal Methods in System Design, 11(2):157-185, 1997.

\bibitem{Jeannet}  B. Jeannet and A. Miné: Apron: A Library of Numerical Abstract Domains for Static Analysis. International Conference on Computer-Aided Verification. LNCS, vol. 5643, pp. 661-667. Springer, Heidelberg, 2009.


\bibitem{Putot} E. Goubault and S. Putot, Static Analysis of Numerical Algorithms. International Static Analysis Symposium. LNCS, vol. 4134, pp. 18-34. Springer, Heidelberg, 2006.


\bibitem{knapik} M. Knapik, and W. Penczek. Bounded model checking for parametric timed automata. Transactions on Petri Nets and Other Models of Concurrency V. Springer Berlin Heidelberg, pp. 141-159, 2012.

\bibitem{Larrosa} J. Larrosa, T. Schiex. Solving weighted CSP by maintaining arc consistency. Artificial Intelligence, Volume 159, Issue 1, 2004, pp. 1-26, ISSN 0004-3702.


\bibitem{Lecoutre} C. Lecoutre, F. Boussemart, and F. Hemery. Exploiting Multidirectionality in Coarse-Grained arc consistency algorithms. In Proceedings CP'03, pp. 480-494, Kinsale, Ireland, 2003. 


\bibitem{Logozzo} F. Logozzo and M. F\"{a}hndrich. Pentagons: A weakly relational abstract domain for the efficient validation of array accesses. Science of Compututer Programming Journal, 75(9), pp. 796-807, 2010.


\bibitem{Mackworth} A.K. Mackworth. Consistency in networks of relations. Artificial Intelligence, 8, 1977, 99-118.


\bibitem{AMine} A. Miné. A New Numerical Abstract Domain Based on Difference Bound Matrices. Programs as Data Objects II, vol. 2053 of LNCS. pp. 155-172. 2001.

\bibitem{Mine} A. Miné. The octagon abstract domain. In Proc. of Analysis, Slicing and Tranformation. In Working Conference on Reverse Engineering, pp. 310-319. IEEE CS Press, 2001.

\bibitem{Motzkin} T. S. Motzkin. Beiträge zur Theorie der linearen Ungleichungen. Doctoral Thesis, University of Basel, 1933. [English translation: Contributions to the theory of linear inequalities, RAND Corporation Translation 22, by D. R. Fulkerson, 1952]. 



\bibitem{Peron} M. Péron and N. Halbwachs. An abstract domain extending difference-bound matrices with disequality constraints. In VMCAI, volume 4349 of LNCS, pp. 268-282. Springer, 2007.



\bibitem{Porschen}  S. Porschen, E. Speckenmeyer, X. Zhao. Linear CNF formulas and satisfiability. Discrete Applied Mathematics, Volume 157, Issue 5, 6 March 2009. 


\bibitem{Sankaranarayanan} S. Sankaranarayanan, H. Sipma, and Z. Manna. Scalable analysis of linear systems using mathematical programming. In Proc. of International Conference on Verification, Model Checking and Abstract Interpretation, number 3385 in Lecture Notes in Computer Science, pp. 21-47. Springer-Verlag, 2005.


\bibitem{Schaefer} T. Schaefer. The complexity of satisfiability problems. In Proceedings 10th ACM Symposium on Theory of Computing (STOC'78), pp. 216-226, 1978.



\bibitem{Sqalli} M. H. Sqalli, L. Purvis and E. C. Freuder. Survey of Applications Integrating Constraint  Satisfaction  and  Case-Based Reasoning. The  First  International  Conference  and Exhibition on the Practical Application of Constraint  Technologies  and  Logic  Programming, 1921.


\bibitem{Cohen} R. Upadrasta and A. Cohen.  Potential and Challenges of Two-Variable-Per-Inequality Sub-Polyhedral Compilation. In First International Workshop on Polyhedral Compilation Techniques (IMPACT'11), in conjunction with CGO'11, Chamonix, France, April 2011.


\bibitem{Wallace} R. Wallace, Practical applications of constraint programming. Constraints, Vol. 1. pp. 139-168, 1996.


\bibitem {Zhang}  C. Zhang, Q. Lin, L. Gao, X. Li. Backtracking Search Algorithm with three constraint handling methods for constrained optimization problems, Expert Systems with Applications, Volume 42, Issue 21, 30 November 2015. pp. 7831-7845, ISSN 0957-4174. 


\bibitem{Zhou} D. Zhou, J. Huang, and B. Schölkopf. Learning with hypergraphs: Clustering, classification, and embedding. Advances in neural information processing systems, 2006.



\end{thebibliography}
\end{document}